# Backward lasing of singly ionized nitrogen ions pumped by femtosecond laser pulses

Xiang Zhang,[1] Rostyslav Danylo,[1,2] Zhengquan Fan[1], Pengji Ding,[2,3] Chenhao Kou[1], Aurélien Houard,[2] Vladimir Tikhonchuk,[4] André Mysyrowicz,[2] Yi Liu,[1,2,*]

[1]*Shanghai Key Lab of Modern Optical System, University of Shanghai for Science and Technology, 516, Jungong Road, 200093 Shanghai, China*

[2]*Laboratoire d'Optique Appliquée, ENSTA ParisTech, CNRS, Ecole polytechnique, Université Paris-Saclay, 828 Boulevard des Maréchaux, 91762 Palaiseau cedex, France*

[3]*Division of Combustion Physics, Department of Physics, Lund University, P.O. Box 118, SE-221 00 Lund, Sweden*

[4]*Centre Lasers Intenses et Applications, University of Bordeaux-CNRS-CEA, 351 Cours de la Liberation, 33405 Talence cedex, France*

[*]*Corresponding author: yi.liu@usst.edu.cn*

## Abstract

We report on the observation of backward lasing at 391.4 nm of nitrogen ions pumped by linearly polarized intense femtosecond pulses at 800 nm. The strongly enhanced spectral intensity at 391.4 nm, as well as the amplification of an externally injected backward seeding pulse, confirm that the backward 391.4 nm signal is due to optical amplification in the gas plasma. Compared to the forward emission at 391.4 nm, the optimal backward emission is achieved at a lower gas pressure around 10 mbar, which is due to asymmetry of the backward and forward directions rooted in the traveling excitation geometry. This method, using the widely available 800 nm femtosecond pulses as a pump laser, provides a promising scheme for the generation of backward "air laser", which holds a unique potential for optical remote sensing.

Cavity-free lasing of the air molecules or their constituent atoms or ions in a plasma pumped by ultrafast laser is intensively studied in recent 5 years [1-18]. In particular, lasing emission in the backward direction (opposite direction of the pump laser propagation) has attracted much attention [1, 10-18], since it holds unique potential for optical remote sensing applications. In the traditional optical remote sensing, laser pulses from the ground laser station are shoot into the sky and the backward scattered photons or fluorescence signal are detected by the ground observer. Since these emissions are incoherent, emitted in $4\pi$ solid angle, the optical signal collected by the ground observer is very weak [19]. When backward air laser is used for the detection of trace gas or pollutions in atmosphere, the backward laser beam carries the information of the target under investigation [20]. Compared to the scattered photon or the fluorescence, the backward laser beam is emitted in a very small solid angle and therefore the optical signal can be tremendously increased [10, 12-14]. More importantly, coherent nonlinear optical techniques such as Stimulated Raman Gain/loss can be used for air pollution detection [20], while the common detection techniques based on scattered photons or fluorescence are normally incoherent.

Up to now, several methods for generation of backward lasing action with air constituents have been demonstrated. In the first report of Dogariu *et al*, both backward and forward stimulated emissions of oxygen atoms at 845 nm have been observed when ambient air was pumped by picosecond pulses at 226 nm [1]. Later, it was shown that the threshold of backward lasing can be significantly reduced by pre-disassociation of the oxygen molecules with a ns pre-pulse [12]. Recently, it has been reported that nitrogen atoms can lase in a similar manner when air is pumped by intense 206 nm pulses [10]. However, applications of this technique for remote sensing are limited due to a poor transmission of the deep-UV pump pulses in atmosphere. Backward lasing emission of neutral nitrogen molecules was reported in two schemes [13-18]. Strong backward emissions at 337 and 357 nm have been observed when the mixture of a nitrogen and a high pressure argon gas was pumped by linearly polarized mid-infrared pulses at 3.9 or 1.03 μm [13]. However, this technique cannot be used in ambient air since a high-pressure argon (> 3 bar) is necessary for the realization of population inversion [13]. Another observation of backward lasing at 337 nm of neutral nitrogen molecules was reported by some of the current authors when nitrogen gas was pumped with circularly polarized 800 nm femtosecond pulses [14-16]. Unfortunately, the presence of oxygen molecules with the percentage above 13% quenches the backward emission at 337 nm [14]. Therefore, there exists a strong demand for other methods for backward lasing in air pumped by optical pulses in a cavity-free manner.

In this paper, we report on backward lasing at 391.4 nm from singly ionized nitrogen molecules, in addition to the previously observed superradiance in the forward direction [21, 22]. This backward lasing signal presents a sensitive dependence on the pump polarization state, with its intensity dropping down by a factor of 100 if the pump laser ellipticity exceeds 0.3. It is found that for linearly polarized pump pulses at 800 nm, the intensity of the backward 391.4 nm signal can be one order of magnitude higher than its neighboring spectral line at 428 nm, while in the sideway fluorescence the latter is two times more intense. This is a clear demonstration of the optical amplification at 391.4 nm in the backward direction. The optical gain is further confirmed by 10 times amplification of an external seeding pulse injected in the backward direction. These observations suggest a new method for the generation of backward lasing in air, which is intensively searched for applications in remote sensing.

In the experiments, femtosecond laser pulses (800 nm, 12 mJ, 1 kHz) delivered by a commercial laser system (Elite DUO, Coherent Co. LTD) were focused by a convex lens ($f = 300$ mm) into a gas chamber filled with air or pure nitrogen at varying pressure, see Fig. 1. A bright plasma channel of a few millimeters was formed, with its length dependent on the pulse energy and gas pressure. The backward emission from the plasma was collected by a fused silica lens ($f = 100$ mm) behind the dichromatic mirror, which reflects the 800 nm femtosecond pump pulse and transmits the backward emission below 450 nm. The backward emission was send into a fiber and analyzed by a spectrometer. The forward emission from the plasma was also detected with the same fiber spectrometer. To filter out the intense pump pulse and the white light generated due to strong nonlinear interaction, two short pass dichromatic mirrors (reflective for wavelength below 450 nm) were employed. Glass filters (BG 40) were also used to further reduce the residual emission at the fundamental wavelength.

We present in Fig. 2 the spectrum of the sideway fluorescence, forward, and backward emission for the nitrogen pressure of 11 mbar, with 2.15 mJ pump laser pulse energy. The sideway fluorescence consists of many spectral lines, corresponding to the second positive band ($C^3\Pi_u^+$ to $B^3\Pi_g^+$) of the neutral nitrogen molecules and the first negative band ($B^2\Sigma_u^+$ to $X^2\Sigma_g^+$) of the singly ionized nitrogen ions with different vibrational quantum numbers denoted by ν´ and ν. These optical transitions are identified in Fig. 2(a). Strong forward emissions centered at 391.4 and 388.5 nm in Fig. 2 (b) correspond to the P and R branches of the $B^2\Sigma_u^+$ to $X^2\Sigma_g^+$ transition of nitrogen ions. This observation agrees with previous reports of nitrogen ions lasing [4, 21], which was identified as superradiance [21, 22]. The backward emission shown in Fig. 2(c) consists of a strong line at 391.4 nm, while other spectral lines at 337, 357, and 427.8 nm are barely observed. This is the first experimental demonstration of backward lasing of nitrogen ions. It is important to note that the ratio of R and P branches of the backward emission (Fig. 2 (c)) is different from that of the forward emission (Fig. 2 (b)), confirming that the backward signal is not due to the unwanted reflection of the forward lasing beam on the exit window of the gas chamber.

To verify that the nitrogen ions indeed give rise to optical amplification in the backward direction, we have performed a pump-probe experiment with a 100 Hz femtosecond laser system. Similar to the setup in ref. 24, a weak probe beam with central wavelength at 390 nm generated in a BBO crystal was injected in the opposite direction of the pump laser. The pump pulse duration was 40 fs and its energy was 3 mJ. The pump was focused by a convex lens $f$ = 400 mm in a pure nitrogen at 30 mbar. The backward-propagating seed pulse amplified by a factor of 10 at 391.4 nm is shown in Fig. 3, which demonstrates undoubtedly the optical amplification in the backward direction.

We further measured the gas pressure dependence of the backward 391.4 nm signal in pure nitrogen. The results are presented in Fig. 4, together with those of the forward signal. With the gas pressure increasing both the backward and forward 391.4 nm signals become more intense up to an optimal pressure. The signals decrease at higher pressures. Similar pressure dependence of the forward 391.4 nm lasing signal has been reported in Ref. 4 and it was attributed to nonlinear propagation effects.

In order to get further insight into the gas pressure dependence, we compared the intensity of the backward and forward 391.4 nm emissions to the corresponding sideway fluorescence signal in Fig. 5. There are three particular features that deserve our attention. First, the three signals present a rapid increase in the pressure range below 10 mbar. In this regime, the laser pulse propagates in the linear regime and the laser intensity is almost constant, independent of the gas pressure. As a result, the plasma density and the density of the molecules in the excited states are linearly proportional to the gas pressure. This explains largely the rapid increase of the three curves up to gas pressure close to 10 mbar.

The second feature is that the sideway fluorescence is saturated for gas pressures exceeding 20 mbar, while the forward 391.4 nm signal decreases. The sideway fluorescence signal origins from the excited nitrogen ions in the $B^2\Sigma_u^+$ state, and it therefore reflects the density of the plasma. Based on this observation, we conclude that the plasma density keeps almost constant for pressures above 20 mbar, which indicates that the ionization probability $\rho_{ion}/\rho_{neutral}$ of the neutral nitrogen molecules decreases significantly in this pressure range. Since the ionization probability depends on the local laser intensity, we deduce that the laser intensity in plasma decreases progressively for the pressure range 20-100 mbar due to the plasma defocusing effect. This argumentation explains naturally the fact that the forward lasing signal decreases for pressures above 20 mbar, since the optical gain of the $B^2\Sigma_u^+$ to $X^2\Sigma_g^+$ transition depends on the laser intensity and there exists an intensity threshold for lasing action [21].

The third feature in Fig. 5 is that the optimal gas pressure for backward lasing, around 10 mbar, is smaller than that for forward lasing (20 mbar). This confirms again that the backward signal is not due to the reflection of the forward 391.4 nm lasing beam, since otherwise they should present the same pressure dependence. How should we understand this difference in optimal gas pressure? In a laser amplifier, the output amplified signal in the small signal regime is an exponential function of the length $l$, that is $I = I_0 \exp(gl)$, where $g$ is the optical gain. For the forward emission, the length of the amplifier is equal to the geometrical length of the plasma, since a forward-propagating photon experiences the same optical gain along the plasma. Therefore, the optical gain $g$ increases for the pressure range 4-20 mbar, while the gain length $l$ is almost constant. In contrast, for the photons in the backward direction, the effective length for the optical amplification is determined by the lifetime $\tau_g$ of the optical gain, $l_{back} = c\,\tau_g$ [15, 16], which is shown to decrease for higher pressures [10]. As a result, the product of the optical gain and the effective amplification length, $gl_{back}$, depends on the pressure in a more complex manner. For pressure range between 4-20 mbar, the optical gain is increasing, while the effective length of the gain decreases gradually. This explains why the maximum of the gain length product is achieved at a gas pressure below 20 mbar.

We now discuss the possibility of remote backward lasing in air based on the 391.4 nm emission. At first sight, the prospect looks grim because the lasing efficiency peaks at 10 mbar and diminishes rapidly at higher gas densities. However, we recall that the process of filamentation of a near IR femtosecond laser pulse leads to a local reduction of air molecules along a thin channel defined by the filamentation process [25]. With a high repetition rate laser, one may achieve local air molecule densities in the range of optimum conditions for backward lasing. Therefore, a possible scheme of the air lasing would be the following. First, a low density channel is prepared by high repetition rate filamentation. Then the procedure described in Ref. 26 is used: several plasma channels are formed in alignment. The backward emission from the filament furthest away from the laser source crosses successively several plasma channels that are appropriately delayed in time to insure a maximum optical amplification. While such a scheme looks complex, it might become feasible with a development of the femtosecond laser technology.

In conclusion, we have shown that the singly ionized nitrogen molecules pumped by intense 800 nm femtosecond pulses can give rise to coherent emission at 391.4 nm in the backward direction, in addition to the widely reported forward radiation at the same wavelength. With the current focusing geometry, the optimal nitrogen gas pressure was found to be around 10 mbar, less than that of the forward signal. This is explained by the fact that the effective length of amplification depends on the gas pressure for backward propagating photons. This process may lead to backward lasing in ambient air, assuming that the local air density can be significantly depressed in the core of a preformed femtosecond plasma filament in air.

Acknowledgement

The work is supported by the National Natural Science Foundation of China (Grants No. 11574213), Innovation Program of Shanghai Municipal Education Commission (Grant No. 2017-01-07-00-07-E00007). Y. Liu acknowledges the support by The Program for Professor of Special Appointment (Eastern Scholar) at Shanghai Institutions of Higher Learning (No. TP2014046) and Shanghai Municipal Science and Technology Commission (No. 17060502500).

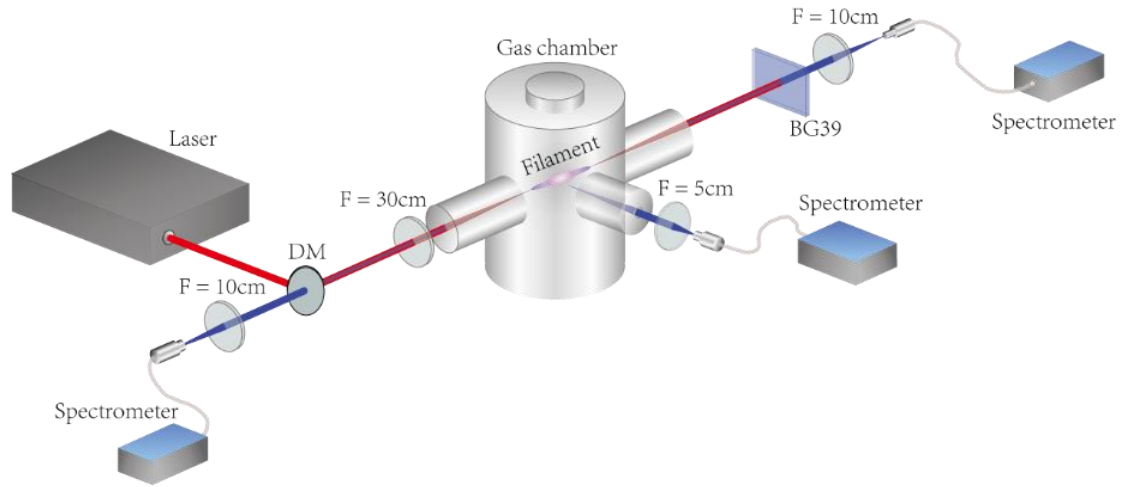

Figure 1. Experimental setup. The femtosecond laser pulses were focused by the $f$ = 300 mm lens into the gas chamber to form a plasma. The emission from the plasma in the backward, forward, and sideway were collected by a fiber connected to a spectrometer for analysis. The short-pass filters (BG40) were used for detection of the backward and forward emissions. DM: dichromatic mirror.

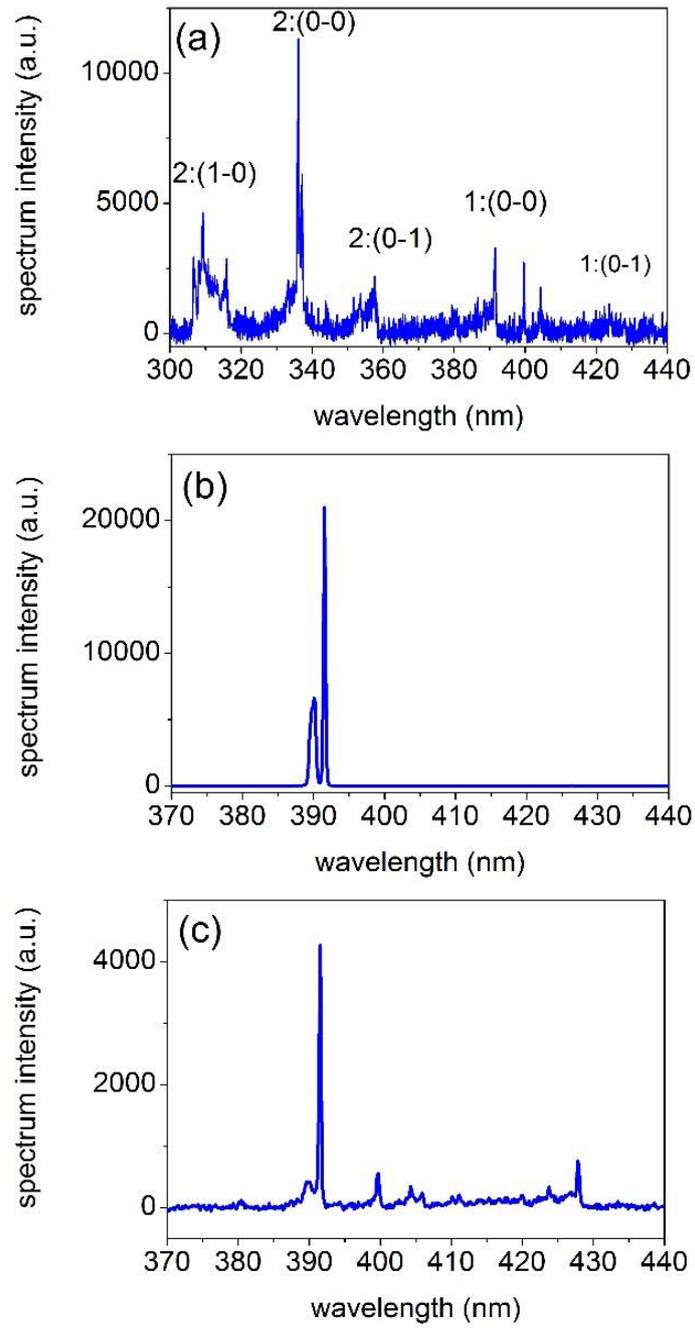

Figure 2. Spectrum of the sideway fluorescence (a), forward (b) and backward (c) signal from the nitrogen plasma. The pump laser energy is 3 mJ and the gas pressure is 11 mbar.

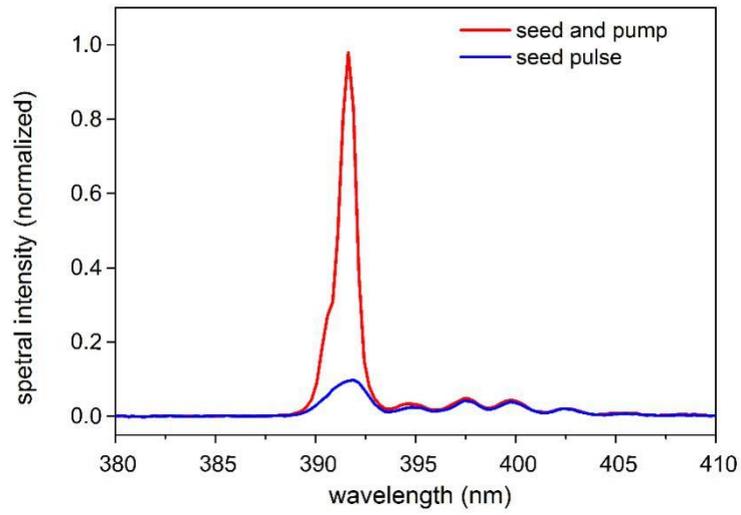

Figure 3. Amplification of the backward-propagating seeding pulse in a plasma. Blue line: seeding pulse, red line: the amplified emission in presence of both the pump laser and seeding pulse. The energy of the pump laser is 7.5 mJ, it was focused in 20 mbar nitrogen gas with a convex lens $f$ = 400 mm.

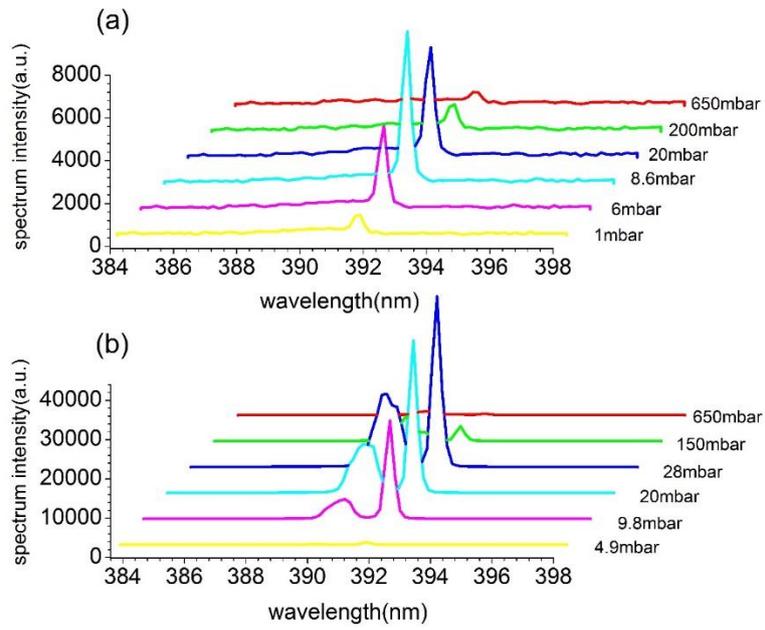

Figure 4. Pressure dependence of the backward (a) and forward (b) lasing emission. The pump pulse energy was 3.5 mJ.

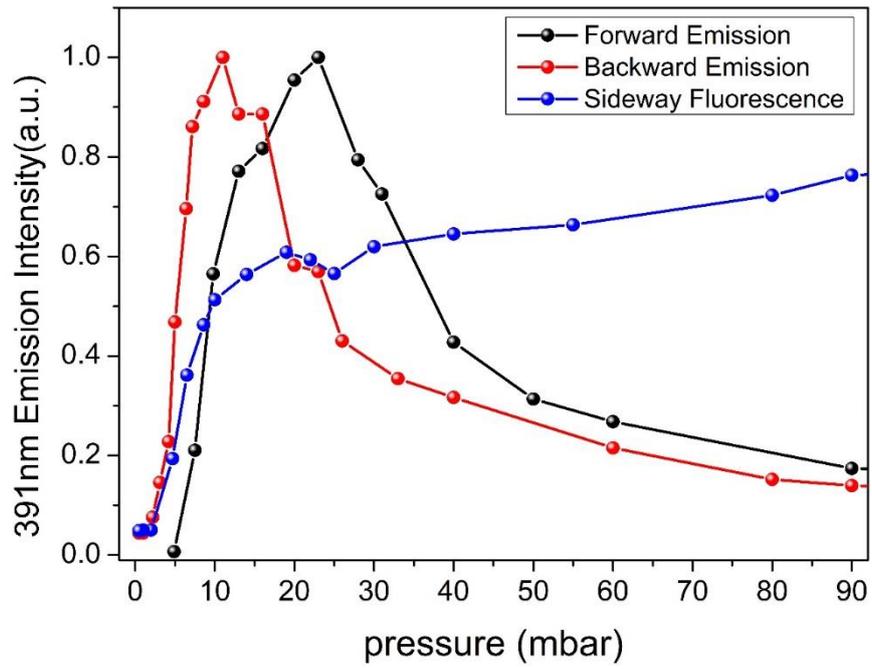

Figure 5. Intensity of the sideway fluorescence at 391. 4 nm (black dot), the forward superradiance signal (blue dot) and the backward 391.4 nm emission (red dot) as a function of the gas pressure. The pump laser energy was 3.5 mJ.